\documentclass[runningheads]{llncs}
\usepackage[T1]{fontenc}
\usepackage{graphicx}
\usepackage{multirow}
\usepackage{arydshln}

\usepackage{float}

\begin{document}
\title{Splitting User Stories Into Tasks with AI - A Foe or an Ally?}
\titlerunning{Splitting User Stories Into Tasks with AI - A Foe or an Ally?}
% If the paper title is too long for the running head, you can set
% an abbreviated paper title here
%
\author{Luka Pavlič\inst{1}\orcidID{0000-0001-5477-4747} \and
Reinhard Bernsteiner\inst{2}\orcidID{0000-0002-8142-3544} \and
Stephan Schlögl\inst{2}\orcidID{0000-0001-7469-4381} \and
Christian Ploder\inst{2}\orcidID{0000-0002-7064-8465}}
\authorrunning{L. Pavlič et al.}
% First names are abbreviated in the running head.
% If there are more than two authors, 'et al.' is used.
%
\institute{University of Maribor, Faculty of Electrical Engineering and Computer Science \\ Koroška cesta 46, 2000 Maribor, Slovenia \\ \and
Department of Management, Communication \& IT, MCI Management Center Innsbruck, 6020 Innsbruck, Austria \\
\email{Luka.Pavlic@um.si, \{ReinhardChristian.Bernsteiner, Stephan.Schloegl, Christian.Ploder\}@mci.edu }}
\maketitle              % typeset the header of the contribution
\begin{abstract}
In agile software development, breaking down user stories into actionable tasks is a critical yet time-consuming process. This paper investigates the potential of Generative AI tools to assist in task splitting, aiming to enhance planning efficiency. We conducted a controlled experiment comparing traditional task-splitting methods with AI-assisted approaches using GitLab Duo. Our findings indicate that while current AI tools are not yet mature enough to replace developers, they can aid in generating more granular task lists and ensuring no important tasks are overlooked. Participants favored a hybrid approach, combining AI tools with conventional methods to maintain high accuracy in planning. This study highlights the potential benefits and limitations of integrating Generative AI into agile development processes, suggesting that AI tools can serve as valuable aids in task splitting, provided there is human oversight to filter out irrelevant tasks. 

\end{abstract}
\section{Introduction}

In software engineering, defining a well-structured set of actionable items - such as activities, tasks, and software components - is crucial for delivering a complete product. In agile development, requirements are typically formulated as user stories, which are then broken down into tasks before implementation. This process usually takes place during planning activities. For instance, in Scrum, the development team creates a sprint backlog during sprint planning, where user stories are split into tasks for the upcoming sprint \cite{13_book_agile_estimation_and_planning}. The industry standard for time allocated to planning is generally around 5\% of the total project time \cite{2_PMBOK2013}. In Scrum \cite{schwaber2017definitive_scrum}, the leading agile development framework \cite{4_agile_report}, planning is conducted in a dedicated session that spans the first half day of every sprint. Given that sprints typically last two weeks (10 working days), this means teams often approach or even exceed the recommended time for planning.

Due to the time-intensive nature of sprint planning, various approaches have emerged to introduce automation into this process. Recent advancements in AI - particularly in Generative AI, which is transforming the software development landscape - suggest that novel tools could enhance planning efficiency.

To explore the feasibility of AI-driven planning, we conducted a controlled experiment. In this study, development teams were given full access to an existing information system and tasked with upgrading it based on well-documented requirements. Before implementation, they planned their activities by breaking down user stories into tasks. The experimental group used a Generative AI-based tool (GitLab Duo) to assist in task splitting, while the control group relied on conventional methods, such as experience-based or analogy-based planning. Empirical data were collected through pre- and post-test questionnaires, as well as observations during the planning and development phases.

One aspect of our research focused on evaluating the potential of Generative AI tools for effort estimation. Our findings \cite{app142412006_pavlic} indicate that current Generative AI-based tools integrated into software development platforms are not yet mature enough to replace human estimators. The average accuracy of automated effort estimation was only 16\%, making it unsuitable for industrial applications \cite{app142412006_pavlic}.

However, preliminary results suggest that Generative AI tools could still be beneficial for assisting in task breakdown. Even in their current form, these tools can aid in generating tasks and ensuring that no important tasks are overlooked during traditional planning. Participants in our study expressed a clear preference for a hybrid approach, combining conventional methods with Generative AI tools to maintain high accuracy in planning.

Building on these insights, we conducted a controlled experiment to evaluate a selected Generative AI- and LLM-based assistant in an iteration-long development cycle. Development teams were asked to split requirements into tasks and implement them. We analyzed empirical data gathered during the experiment to compare traditional task-splitting methods with AI-supported task generation. To achieve our research objective, we address the following research question:

\begin{itemize}
    \item \textbf{RQ:} To what extent can Generative AI-supported tools assist development teams in breaking down user stories into tasks?
\end{itemize}

To answer this question, we examine the similarities and differences between AI-generated and human-created task sets. Additionally, we assess whether AI-generated tasks were appropriate and useful during implementation. The methodology used to investigate this research question is outlined in the “Research Method” section, following the “Related Work” section. After presenting the results of our controlled experiment, we will discuss our findings, outline potential threats to validity, and conclude the paper.

\section{Related Work}

Task splitting is a fundamental process in software development, ensuring efficient workflow and clear responsibility allocation. Traditional methods for task decomposition have been widely explored and applied in industry. Several established approaches exist for decomposing complex projects into manageable tasks \cite{2_PMBOK2013,task_decomposition,human_decompose_tasks,spidr,invest}:

\begin{itemize}
    \item \textbf{Functional Decomposition} is a technique that involves breaking down a system into smaller, manageable functions or modules, where each function represents a specific task or responsibility, facilitating easier development and maintenance.
    \item \textbf{Top-Down} and \textbf{Bottom-Up} Design: In top-down design, a system is decomposed from the highest level into smaller components, prioritizing the overall architecture before detailing specific parts. In contrast, bottom-up design begins with detailed components and integrates them to form a complete system. A hybrid approach is often used to leverage the advantages of both methods.
    \item \textbf{Workflow Steps} method involves analyzing the workflow associated with a user story and splitting it based on distinct steps or stages. By identifying each step required to achieve a goal, development teams can create tasks that correspond to these individual steps, ensuring a comprehensive implementation of the user story.
    \item \textbf{SPIDR Technique}, proposed by Mike Cohn, SPIDR stands for Spike, Path, Interface, Data, and Rules. This technique enables the systematic splitting of large user stories by investigating uncertainties (Spike), considering different execution paths (Path), focusing on interfaces (Interface), varying data inputs (Data), and applying business rules (Rules).
    \item Often employed \textbf{INVEST Criteria} stands for Independent, Negotiable, Valuable, Estimable, Small, and Testable. This guideline ensures that user stories are well-formed and can be effectively split into tasks. By adhering to these criteria, teams can create tasks that are manageable and deliver distinct value.
\end{itemize}

Recent research has explored the application of AI-based approaches, particularly machine learning techniques, in task decomposition. The study “User Story Splitting in Agile Software Development using Machine Learning” leverages machine learning techniques to automate user story splitting, enhancing the efficiency and accuracy of task decomposition.

The thesis “Splitting User Stories Using Supervised Machine Learning” \cite{10053226} introduces a novel approach for automatically splitting user stories into tasks using supervised machine learning algorithms, including random forest and decision trees. The goal is to improve development efficiency by reducing manual effort in task decomposition.

The study “Machine Learning-Based Approach for User Story Clustering in Agile Engineering” \cite{article} applies K-means and K-medoids clustering algorithms to organize user stories into cohesive clusters, facilitating better understanding and implementation at early development stages.

In addition, several studies and tools have recently addressed the use of generative AI for task decomposition. E.g. the research “Automated User Story Generation with Test Case Specification Using Large Language Models” \cite{inproceedings_automated_us_generation} introduces "GeneUS," a tool leveraging GPT-4 to generate user stories from requirements documents, facilitating integration with project management systems and streamlining requirements engineering.

Authors \cite{10.1007/978-3-031-61154-4_8} of “LLM-Based Agents for Automating the Enhancement of User Story Quality” explores the use of large language models to improve user story quality within agile teams. The researchers developed an autonomous LLM-based agent system to assess its effectiveness in enhancing user story quality.

“MASAI: Modular Architecture for Software-Engineering AI Agents” \cite{Arora2024MASAIMA} proposes a modular architecture for AI agents in software engineering, where different LLM-powered sub-agents are instantiated with specific objectives. This approach enhances task decomposition and coordination among agents, improving software development processes.

The potential of generative AI in iteration planning has been highlighted in recent Gartner Research Reports \cite{23_book_gartner}. By 2027, it is projected that the number of platform engineering teams using AI to augment every phase of the software development lifecycle will increase from 5\% to 40\%. Additionally, Gartner’s 2024 report identified state-of-the-art automation tools for planning iterations, including Atlassian Intelligence, ClickUp, GitLab Duo, and OpenText. Based on Gartner’s recommendations and our preliminary tool selection and testing, we decided to use GitLab Duo, which is integrated into GitLab Enterprise. As one of the leading tools, it provides a strong foundation for generalizing our research findings. Our study is unique in that we apply generative AI explicitly for user story splitting, aligning with expert-based estimation techniques, such as planning poker. This distinction enables direct validation of our research and highlights its potential impact on industry practices.

\section{Research Method}

To investigate the research question, we designed a controlled experiment iteratively, refining it based on a pilot study conducted in a small setting. The experiment follows a well-established scoping process used in empirical software engineering research \cite{11_book_experimentation}. Our goal was to evaluate the performance of two competing approaches to task splitting in software development: a conventional method and an AI-assisted approach integrated within a software development platform. The controlled experiment was structured as a one-factor experiment \cite{practical_guide_experimentation}, where the independent variable was the method of task splitting. Participants were divided into two groups:

\begin{itemize}
    \item \textbf{Control Group}: Used conventional task splitting techniques.
    \item \textbf{Experimental Group}: Utilized GitLab Duo, an AI-powered tool, to automatically generate tasks based on user story descriptions.
\end{itemize}

We recruited higher-year computer science students, most of whom had industry experience as software developers. A detailed breakdown of participant demographics is provided in Section Results.
The experiment consisted of three sessions, simulating a sprint-long development:

\begin{itemize}
    \item \textbf{Session 1}: Requirement presentation, pre-test, and participant distribution.
    \item \textbf{Session 2}: Requirement implementation.
    \item \textbf{Session 3}: Acceptance of software increments and post-test questionnaire.
\end{itemize}

\subsection{Session 1: Setup and Pre-Test}

\begin{figure}
\includegraphics[width=\textwidth]{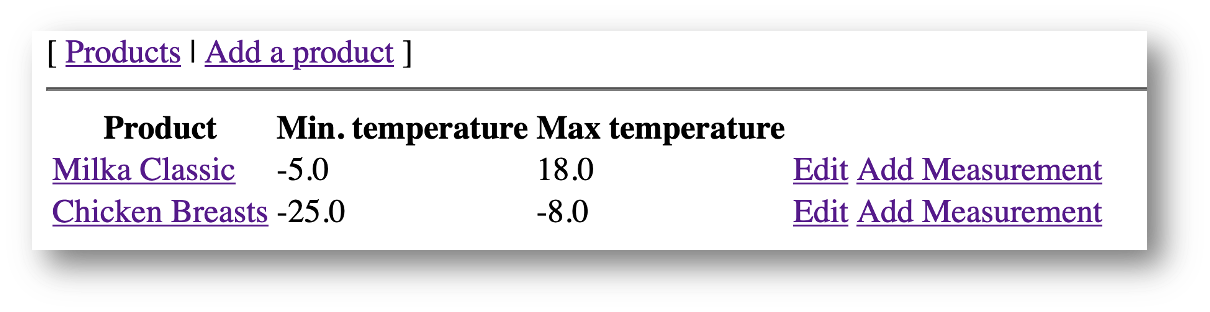}
\caption{Sample user interface of the system that participants were required to upgrade - list of all products.} \label{fig1}
\end{figure}

\begin{figure}
\includegraphics[width=\textwidth]{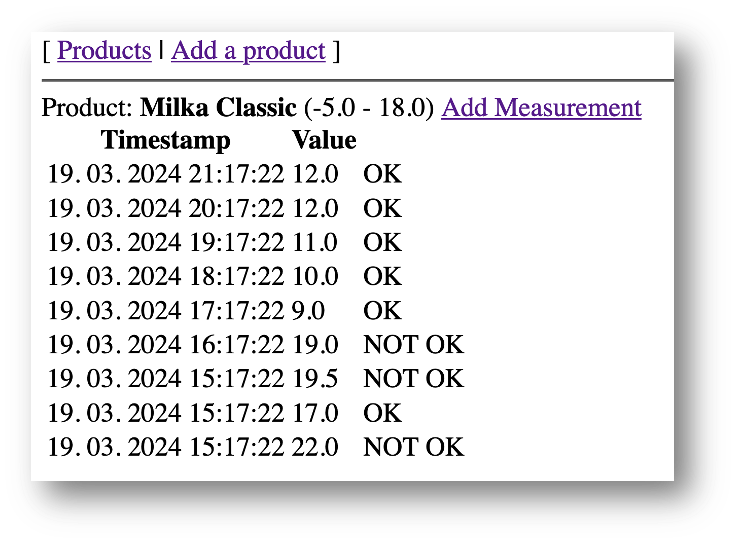}
\caption{Sample user interface of the system that participants were required to upgrade - temperature history data.} \label{fig2}
\end{figure}

The participants were introduced to an existing information system built using Jakarta EE, with a relational database backend and a Jakarta Server Faces-based web interface. The system was designed for monitoring product storage temperatures and alerting deviations (see Figures \ref{fig1} and \ref{fig2}). To ensure that technological unfamiliarity did not bias results, all participants underwent a pre-test questionnaire assessing their experience with the technology stack.

Participants were then randomly assigned to either the control or experimental group, with teams of three simulating agile development settings. Each team received eight user stories, designed to exceed a single iteration’s workload, requiring them to prioritize tasks. The user stories included functionalities such as:

\begin{itemize}
    \item Email notifications for out-of-tolerance measurements.
    \item Monthly reporting of measurement deviations.
    \item Color-coded measurement history.
    \item Support for metric and imperial units.
    \item Graphical representation of product temperature trends.
    \item Importing data from offline sensors.
    \item Temperature trend visualization per product.
    \item Multi-language support (DE/EN/SI).
\end{itemize}

The control group created task sets using their chosen conventional techniques, while the experimental group used GitLab Duo. Participants in the experimental group interacted with the AI assistant by providing prompts to generate task breakdowns. They could refine their prompts but were not allowed to override the AI-generated results.

\subsection{Session 2: Task Implementation}

During this session, teams implemented selected user stories based on their generated tasks. GitLab was used to manage tasks, tracking progress via issue movement on a kanban board.

\subsection{Session 3: Increment Acceptance and Post-Test Questionnaire}

After implementation, user stories were reviewed and tested for correctness. Only successfully implemented user stories were considered for analysis. Participants then completed a post-test questionnaire assessing their perceptions of AI-assisted task splitting.

\section{Results and Discussion}

\subsection{Participant Demographics and Experiment Execution}

The controlled experiment was conducted in April 2024, involving \textbf{42} developers who met the eligibility criteria based on their prior academic performance. All participants had experience with the technologies used in the information system they were tasked with upgrading, including JSF (Jakarta Server Faces), EJB (Jakarta Enterprise Beans), and JPA (Java Persistence API). Their experience levels were as follows:

\begin{itemize}
    \item the majority (\textbf{91\%}) had \textbf{between 1–3 years} of development experience,
    \item 3\% reported up to 5 years and
    \item 6\% had less than 1 year of experience.
\end{itemize}

Participant involvement was voluntary, ensuring motivation and engagement.
A significant proportion (\textbf{82\%}) of participants had \textbf{prior experience with AI tools such as ChatGPT, GitHub Copilot, Claude, and Superhuman}. Notably, \textbf{85\% expressed a willingness to use AI tools} for breaking down user stories into tasks, while 3\% opposed the idea, and 12\% remained undecided.

Participants were randomly assigned to either the experimental or control group. Each group was further divided into teams of three, forming 7 teams per group. However, one team from the experimental group withdrew, resulting in 39 participants being included in the final analysis (6 teams in the experimental group and 7 in the control group).

\subsection{Task Generation Differences}

The experiment aimed to evaluate the differences between human-generated and AI-assisted task generation. The control group had complete autonomy in task creation and estimation, while the experimental group was required to use GitLab Duo. In addition to task breakdown, the experimental group also reported the prompts used for AI-based task generation.
The analysis of task generation revealed notable differences between the two groups (see Table \ref{tab1}):

\begin{itemize}
    \item The experimental group created a total of 260 tasks, averaging 5.4 tasks per user story, whereas the control group created only 184 tasks, averaging 3.2 tasks per user story.
    \item Human-generated tasks were more focused and directly relevant to fulfilling the user stories, while AI-generated tasks included additional tasks such as unit testing, documentation updates, and code refactoring.
    \item Some AI-generated tasks appeared unrelated to specific user stories, such as "Create a new page template and routing."
\end{itemize}

\begin{table}
\caption{The number of generated tasks.}\label{tab1}
\begin{tabular}{|l|c|c|}
\hline
 &  \textbf{Experimental Group} & \textbf{Control Group}\\
\hline
Development teams &  6 & 7\\
\hdashline
User stories per development team &  8 & 8\\
\hdashline
Split user stories &  48 & 56\\
\hdashline
Generated development tasks &  260 & 184\\
\hdashline
Tasks per development team &  43.33 & 26.29\\
\hline
\end{tabular}
\end{table}

To demonstrate the differences between task sets generated manually, and those generated by the AI-enabled tool, we are outlining three user stories that were implemented by the most teams. In Tables \ref{tab2}, \ref{tab3} and \ref{tab4}, we can see typical tasks for each user story. The obvious observation is that manual generated tasks are smaller in volume. However, after analyzing actual implemented tasks, needed to fulfill the user story, it turned out, that development teams implemented on average only \textbf{59\% of generated tasks}, while teams, that created tasks manually, implemented all of them. Additional insight shows a clear difference in tasks content - auto-generated task sets include tasks, that one would miss in manual-created lists (e.g. tasks, related to testing and updating documentation).

\begin{table}
\caption{Email notifications for out-of-tolerance measurements.}\label{tab2}
\begin{tabular}{|p{7cm}|p{7cm}|}
\hline
\textbf{Control Group} & \textbf{Experimental Group} \\
\hline
Create a new email account & Add email dependency and configure MailSender \\
Prepare the GUI & Add error handling \\
Configure email on the server & Update measurement logic to check tolerances \\
Implement the backend for sending & Create email notification service \\
 & Write tests \\
 & Documentation \\
 & Query database to get product details and temperature measurements \\
 & Add interface for configuring email recipient \\
 & Generate email with product details and measurement when out of range \\
 & Configure service on startup \\
\hline
\end{tabular}
\end{table}

\begin{table}
\caption{Monthly reporting of measurement deviations.}\label{tab3}
\begin{tabular}{|p{7cm}|p{7cm}|}
\hline
\textbf{Control Group} & \textbf{Experimental Group} \\
\hline
Display on the frontend & Create new page template and route \\
Develop the backend (methods, queries) & Add new "Monthly Report" menu option \\
Filter by drop-down list & Add interface for configuring report parameters \\
Display the number of non-compliant measurements for individual products & Aggregate data and populate model \\
 & Update data aggregation job \\
 & Query database for measurements from last 30 days \\
 & Implement service to query measurements and generate report \\
 & Group report data by product \\
 & Display data on new page \\
 & Unit tests for report generation logic \\
 & Integration tests for end-to-end report generation \\
 & Call monthly report service on menu option selection \\
\hline
\end{tabular}
\end{table}

\begin{table}
\caption{Color-coded measurement history.}\label{tab4}
\begin{tabular}{|p{7cm}|p{7cm}|}
\hline
\textbf{Control Group} & \textbf{Experimental Group} \\
\hline
Cell coloring regarding measurement value & Write CSS rules to color adequate measurement rows green and inadequate rows red \\
 & Implement conditional color logic \\
 & Write unit tests for the updated MeasurementHistory component \\
 & Update documentation explaining the color thresholds and styling \\
\hline
\end{tabular}
\end{table}

These findings suggest that AI tools can generate a more granular task list, which could be useful in highlighting overlooked steps. However, the AI-generated tasks also introduced potentially irrelevant activities, reinforcing the need for human oversight.

\subsection{Final Remarks}

In addition to assessing task generation differences, the study investigated participants’ perspectives on AI-assisted task breakdown. The post-test questionnaire revealed key insights:

\begin{itemize}
    \item When asked about their preferred approach for future projects, all participants (100\%) \textbf{favored a combination of AI tools and conventional methods}. None preferred exclusively AI-supported or conventional methods.
    \item Regarding AI tools' accuracy, \textbf{only 10\% believed AI tools provided more relevant tasks} than conventional methods, while 55\% disagreed, and 35\% were uncertain.
\end{itemize}

The empirical data supports the conclusion that AI tools can assist in task generation but should not be relied upon exclusively. The control group’s task lists were more contextually relevant, while the AI-generated lists included a broader scope of activities, some of which were not explicitly required.
The experiment demonstrates that AI tools can enhance task generation by increasing granularity and identifying overlooked tasks.

\section{Threats of Validity}

Our experiment reflects the current state of generative AI, but rapid advancements may soon render our findings outdated. We used a leading AI tool in its general form, though fine-tuning for a specific domain or incorporating feedback loops could improve accuracy. The experimental group’s estimates lacked developer-specific pre-knowledge, unlike the control group, affecting comparability. While participants were well-trained software engineering students, their lack of full-time industry experience may limit generalizability. The controlled setting ensured consistency but involved smaller-scale tasks and a shorter development phase than typical projects. Lastly, restricting the study to GitLab Duo, while justified by its industry standing, may limit the applicability of results to other AI-powered tools.

\section{Conclusions}

Task splitting is one of critical activities in agile software development. While traditionally time-consuming, recent advances in AI, especially generative AI, promise to streamline this process. To test these advancements, we conducted a controlled experiment to evaluate AI tools in task splitting during development.

Our research found that current AI-based task splitting helper tools, like GitLab Duo, are not yet reliable replacements for human developers. However, they show potential as helpful aids in the task splitting process, offering valuable assistance in identifying overlooked tasks or verifying that all necessary tasks are included during iteration planning.

Our analysis revealed two key areas affected by AI-supported task splitting: identifying missing tasks that human assessors often forget, additional context-ignorant tasks generated automatically. In conclusion, generative AI tools can serve as valuable aids in the task splitting phase, helping ensure more comprehensive and accurate task generation.

\begin{credits}
\subsubsection{\ackname} The authors acknowledge the financial support from the Slovenian Research Agency (Research Core Funding No. P2-0057). 
\end{credits}

\bibliographystyle{splncs04}
\bibliography{references.bib}

\end{document}